\documentclass{article}

\usepackage{graphicx} 
\usepackage{amsmath}
\usepackage{mathtools}
\usepackage[caption = false]{subfig}
\usepackage[section]{placeins}
\usepackage{xcolor}

\usepackage{algpseudocode}
\usepackage{algorithm}

\usepackage{hyperref}
\hypersetup{
    colorlinks=true,
    linkcolor=blue,
    filecolor=magenta,      
    urlcolor=cyan,
    pdftitle={Overleaf Example},
    pdfpagemode=FullScreen,
}

\usepackage[
	backend=bibtex,
	sorting=ynt
]{biblatex}

\newcommand{\X}{\textbf{X}}

\newcommand{\N}{\text{N}}
\newcommand{\xrow}{\boldsymbol{x}_i'}
\newcommand{\xrowr}{\boldsymbol{x}_{il}'}

\title{False Discovery Rate Control via Bayesian Mirror Statistic}
\author{Marco Molinari \\ University of Oslo \and Magne Thoresen \\ University of Oslo}
\date{October 2025}

\begin{document}
\maketitle

\begin{abstract}
Simultaneously performing variable selection and inference in high-dimensional models is an open challenge in statistics and machine learning. The increasing availability of vast amounts of variables requires the adoption of specific statistical procedures to accurately select the most important predictors in a high-dimensional space, while being able to control some form of selection error. In this work we adapt the Mirror Statistic approach to False Discovery Rate (FDR) control into a Bayesian modelling framework. The Mirror Statistic, developed in the classic frequentist statistical framework, is a flexible method to control FDR, which only requires mild model assumptions, but requires two sets of independent regression coefficient estimates, usually obtained after splitting the original dataset. Here we propose to rely on a Bayesian formulation of the model and use the posterior distributions of the coefficients of interest to build the Mirror Statistic and effectively control the FDR without the need to split the data. Moreover, the method is very flexible since it can be used with continuous and discrete outcomes and more complex predictors, such as with mixed models. We keep the approach scalable to high-dimensions by relying on Automatic Differentiation Variational Inference and fully continuous prior choices.
\end{abstract}

\section{Introduction}\label{sec:Intro}
Advances in data collection capabilities have allowed researchers to get access to thousands of features on multiple subjects in relatively short times. A typical example is biomarker discovery, where thanks to next-generation DNA and RNA sequencing technologies, the number of variables is usually much higher than the sample size (\cite{Berry2020}). Moreover, in particular in clinical settings, measurements are often taken multiple times, over a given time-frame and for diverse interventions. In these highly complex designs, feature selection becomes a challenging and critical task, where error control is key to limit the number of false discoveries. The combination of these aspects provides the motivation for this work, where we aim at providing a way to control the False Discovery Rate (FDR, \cite{Benjamini1995}) for a wide range of models, for example in the presence of repeated measurements, interactions or non-normally distributed outcomes.\\

Several methods constructed for variable selection in high-dimensional models already exist in literature, some popular choices being the LASSO, Elastic Net, LARS and SCAD, which provide efficient algorithms that scale well to a large number of features (\cite{Tibshirani1996, Zou2005, Efron2004, Fan2011}). However, inference on the parameters of interest and FDR control is more cumbersome and not always possible. Simply proceeding with inference (for example through OLS), after a data-dependent variable selection step (such as the ones produced by LASSO and others), does not allow to perform valid \textit{post-selection} inference (\cite{Berk2013}); confidence intervals will be biased, leading to a potential increase in false discoveries. To this end, several methods for FDR control exist, each with strengths and limitations. The \textit{Mirror Statistic} of \cite{Xing2021}, later adopted in conjunction with data splitting by \cite{Dai2022}, is a powerful method to perform FDR control without requiring many assumptions about the data generating mechanism. However, although Mirror Statistic with data splitting is very general, the split operation can drastically reduce the power of the model and the ability to recover the set of true active variables. This is where we move to a Bayesian approach, with the aim of combining the strengths of Bayesian inference and the Mirror Statistic theory.\\
In the Bayesian framework, we take a fundamentally different view on the nature of the parameters, assuming they are unknown random variables, rather than unknown scalar values. The aim of Bayesian inference is then to estimate these unknown distributions. This implies that once we have estimated our model, we will also have an estimate of the variability  of the parameters, and it is this key additional information that we exploit to adapt the Mirror Statistic into a Bayesian framework, so that we can construct the mirror coefficients with a single run of the model, without the need to split or randomise the data.\\
Choosing the prior distribution plays a fundamental role in controlling the level of shrinkage that we want to impose on the coefficients of the model and in the context of high-dimensional data, it is vital to choose a prior that can effectively shrink coefficients to $0$. Classic examples are the Spike and Slab and the Horseshoe, see \cite{OHara2009} and \cite{Schoot2021}, among others, for a review of some popular distributions.\\
FDR control has long been used also with Bayesian models, see for example \cite{Efron2001, Newton2004, Muller2006FDR, Efron2008, Stephens2016} for an in-depth analysis of some existing approaches as well as an analysis of the conceptual differences between the frequentist and the Bayesian view of FDR. These approaches rely on the specification of a \textit{two-groups} model, i.e., a mixture distribution (like a Spike and Slab) as prior on the regression coefficients, which then allows to threshold the posterior inclusion probabilities to control FDR. This has the disadvantage that the model definition is limited by the class of prior distributions that provide such probabilities and since the prior includes discrete components, inference is more cumbersome, especially in high-dimension. \cite{Buettner2021} propose another approach that does not use a mixture prior but thresholds regression coefficients using an arbitrary cutoff, in order to perform FDR control.\\
In our proposed strategy we want to build a model where all prior distributions are continuous, which means avoiding mixture distributions. We only require that the prior of choice imposes shrinkage on the regression coefficients towards $0$. We want to do so in order to retain a simpler and computationally efficient model specification and being able to use automatic differentiation software to perform inference. This combination of factors gave us the motivation to use the concept of the Mirror Statistic as a way to control FDR in a wider range of Bayesian models.\\

The rest of the paper is organized as follows: in Section \ref{sec:FDR_control} we review the concept of FDR and Mirror Statistic, and we introduce our contribution, the Bayesian version of the Mirror Statistic. In Section \ref{sec:model} we provide the details about the model and prior distributions that we use. In Section \ref{sec:simulations} we show the performance of the proposed strategy on several simulations and in Section \ref{sec:real_data} we use it on a real dataset. Finally, in Section \ref{sec:Discussion} we summarise our contribution and provide a discussion on limitations and potential improvements.

\section{False Discovery Rate control}\label{sec:FDR_control}
Throughout the section we refer to the general problem of variable selection and FDR control for a general regression model $y_i = f(\sum_{j=1}^p x_{ij} \beta_j) $.\\
We denote the available data with $\text{D}_i = \{y_i, x_{i1}, \dots, x_{ip}\}$; a collection of $n$ samples of outcome $y_i$ and $p$ variables $x_j$. Furthermore, let $r_j \in \{0, 1\}$ denote the unknown ground truth indicator for the presence/absence of the effect $\beta_j$ and let $\delta_j \in \{0, 1\}$ be the binary decision function for the inclusion/exclusion of the effect $\beta_j$.\\
Using this notation, the False Discovery Proportion (FDP) is a random variable defined as the proportion of falsely included covariates over the total number of included ones
\begin{align}\label{eq:FDP}
    \text{FDP} = \dfrac{ \sum_{j=1}^{p} \delta_j (1 - r_j) }{ \sum_{j=1}^{p} \delta_j \vee 1 }
\end{align}
The False Discovery Rate is then generally defined as the expected value of the FDP, $\text{FDR} = \text{E}\left[ \text{FDP} \right]$, and the aim is to have it to equal to a target level $\alpha \in (0, 1)$.\\
The difference in FDR control between the frequentist and Bayesian approach is the expected value that they aim to control.\\

\subsection{Frequentist FDR control}
In the frequentist framework the FDR is defined as the expectation of the FDP over repeated experiments, $\text{FDR} = \text{E}_D\left[ \text{FDP} \right]$, where $D$ here represents the data generating mechanism. \cite{Benjamini1995}, in their seminal paper, provide a way to control FDR by using estimated $p$-values, under the assumption of independence. \cite{Benjamini2001} later extended the method to work also with positively-dependent $p$-values.\\
The reliance on $p$-values can however be quite restrictive since many algorithms, especially in high-dimensional settings, do not provide $p$-values at all, such as the LASSO (\cite{Tibshirani1996}). In their pioneering work, \cite{Barber2015} introduce the \textit{knockoff} method, as a new way to control FDR that does not have this limitation. Their original development provides a method to control FDR in low-dimensional regression models ($p < n$) when the joint distribution of the covariates $X$ is known.\\
The knockoff method works by performing variable selection on the parametric space augmented by the knockoff version of the original covariates, $X^{(k)}$. $X^{(k)}$ have to be generated such that the new variables have the same dependence structure as $X$, while being conditionally independent of the outcome, and exchangeable. Variable selection is then performed using a test statistic $W_j$, constructed to depend on $y$, $X$ and $X^{(k)}$, i.e. $W_j = g(X, X^{(k)}, y)$, for some function $g$. This test statistic must satisfy the \textit{anti-symmetry} property (\cite{Barber2015}), which means that swapping a variable with its knockoff version will change the sign of $W_j$. For example, given the LASSO estimates of a regression coefficient, $\hat{\beta}_j$, and its knockoff counterpart, $\hat{\beta}_{j}^{(k)}$, $w_j$ can be calculated as $w_j = |\hat{\beta}_j| - |\hat{\beta}_{j}^{(k)}|$, where a large positive value of $w_j$ provides evidence that $y$ depends on $X_j$.\\
Under the assumption that, when $r_j = 0$, the sampling distribution of at least one of the two coefficients, $\beta_j$ and $\beta_{j}^{(k)}$, is symmetric around zero, then also $W_j$ will be symmetric around zero. Using this construction, the authors provide the fundamental new result on how to estimate an upper bound on the number of false positives, defined as follows:
\begin{align}\label{eq:FP_Upper_Bound}
    \sum_{j=1}^{p} \mathbf{1}(w_j > t)(1 - r_j) 
    \approx \sum_{j=1}^{p} \mathbf{1}(w_j < -t)(1 - r_j)
    \leq \sum_{j=1}^{p} \mathbf{1}(w_j < -t), \forall t > 0
\end{align}
where $\mathbf{1}(w_j < -t)$ is a binary indicator function taking the value $1$ when the inner condition is satisfied and $0$ otherwise. Starting from the left-hand side we have the actual unknown number of false positives, which, under the symmetry assumption and the mirroring transformation $g$, is approximately equivalent to the middle term, which is then bounded by the sum on the right-hand side. This term does not include the unknown ground truth $r_j$ and can therefor be used to effectively approximate the numerator of the FDP in Equation \ref{eq:FDP}\\
\cite{Candes2018} extended the methodology to the high-dimensional case ($p > n$); however, the generation of the knockoff variables remains limited by the strong requirement of knowing the joint distribution of the covariates, an information rarely known in practice. To overcome this limitation, \cite{Xing2021} propose the \textit{Gaussian Mirrors}, a method based on a new test statistic built on two perturbed versions of each covariate $x_j$, rather than the knockoff variables. They generate the new variables $x_j^{(a)}$ and $x_j^{(b)}$ by carefully adding the right amount of Gaussian noise to $x_j$, so that the two new covariates are independent, but still retaining the dependence with $y$. They then estimate two regression coefficients, $\beta_j^{(a)}$ and $\beta_j^{(b)}$, and compute the new test statistic, called the \textit{Mirror statistic}, using these pairs of coefficients, as follows:
\begin{align*}
    W_j = |\beta_j^{(a)} + \beta_j^{(b)}| - |\beta_j^{(a)} - \beta_j^{(b)}|
\end{align*}
The Mirror statistic $W_j$ has two parts, the first $|\beta_j^{(a)} + \beta_j^{(b)}|$,  accounts for the strength of the signal, while the second, $|\beta_j^{(a)} - \beta_j^{(b)}|$, captures the noise of the estimates (or the variability). This construction is motivated by the fact that, for $r_j = 0$, i.e. when the $j_{th}$ covariate is not active, the estimates of $\beta_j^{(a)}$ and $\beta_j^{(b)}$ will vary around $0$ and the variation will be due to noise only, therefor $W_j$ will also center at $0$. Viceversa, when $r_j = 1$, i.e. $x_j$ is an active covariate, the first component, $|\beta_j^{(a)} + \beta_j^{(b)}|$, will reflect the signal and will be "significantly" far away from $0$, while the second part, $|\beta_j^{(a)} - \beta_j^{(b)}|$, will reflect the variability in the estimate of the two coefficients.\\
Given this construction, the upper bound approximation in Equation \ref{eq:FP_Upper_Bound} is still valid and can be used to control the FDR. Given a target FDR value $\alpha$, the optimal inclusion threshold $t_{\alpha}$ is found by optimizing the following loss:
\begin{align}\label{eq:opt_t}
    t_{\alpha} &= \text{min}\left\{ t>0 : \text{FDP}(t) = \dfrac{\sum_{j=1}^{p} \mathbf{1}(w_j < -t)}{\sum_{j=1}^{p} \mathbf{1}(w_j > t) \vee 1 } \leq \alpha \right\}
\end{align}
This strategy represents an improvement, as it does not rely on knowing the covariance of $X$, but it is still limited in being restricted to continuous covariates and computationally inefficient since only a single covariate at a time can be randomised, therefore the whole process has to be repeated $p$ times.\\
\cite{Dai2022} provides an alternative construction of the Mirror statistic based on data splitting to create two independent sets of observations and obtain two independent estimates of the regression coefficients. This approach, which we from now on will refer to as Mirror Statistic Data Splitting (DS), is very general since it does not depend on a specific outcome or covariate distribution, and it is more efficient than the Gaussian Mirrors, as it operates on all covariates simultaneously. However, in practice, the applicability of DS is limited by the available sample size in high-dimensional studies, which makes the data splitting step too costly in terms of loss of power. Also, to use the false positive upper bound approximation in Equation \ref{eq:FP_Upper_Bound}, the regression coefficients distribution must satisfy the symmetry requirement, which is not achieved, in high-dimensions, when deviating from standard linear regression (\cite{Dai2023}).\\
These limitations gave us the idea of using the results based on the Mirror statistic  and the previous FDR approximation result, in a Bayesian framework, leveraging the natural estimation of a full probability distribution provided by Bayesian inference.

\subsection{Bayesian FDR control}\label{sec:Bayesian_MS}
Bayesian approaches to FDR have been studied by several authors, in particular starting with the work of \cite{Efron2001}, who developed the concept of \textit{local} FDR (commonly denoted as \textit{fdr}). The framework, also used in many following works, is to adopt a \textit{two-groups} model, defined as
\begin{align}\label{eq:mixture_model}
    \beta_j \sim \pi_0 f_0(\beta_j) + (1 - \pi_0) f_1(\beta_j) \equiv f(\beta_j)
\end{align}
where $\pi_0$ is the unknown proportion of \textit{true-null} coefficients, $f_0$ is the coefficient distribution under the null hypothesis and $f_1$ is the distribution under the alternative hypothesis. The local fdr is then defined as
\begin{align}\label{eq:local_fdr}
    \text{fdr}_j = \dfrac{\pi_0 f_0(\beta_j)}{f(\beta_j)} 
    = p(r_j = 0 \mid \hat{\boldsymbol{\beta}}, f_0, f_1, \pi_0, D)
\end{align}
which in a Bayesian context is the posterior probability of effect $j$ being null, given that it was estimated to be non-null.\\
\cite{Newton2004} and \cite{Muller2006FDR} then provide a definition of Bayesian FDR (BFDR) as the expectation of FDP with respect to the model $f$, conditioned on the observed data
\begin{align*}
    \text{BFDR} = \text{E}_{f} [ \text{FDP} \mid \text{D}] = 
    \text{E}_{f} \left[ \dfrac{ \sum_{j=1}^p \delta_j(t)(1 - \nu_j) }{ \sum_{j=1}^p \delta_j(t) } \right]
\end{align*}
where $\nu_j = p(r_j = 1 \mid \text{D})$ is the posterior probability of effect $j$ being non-null, and $1 - \nu_j$ is the individual probability of false discovery, i.e. the local \textit{local} fdr from Equation \ref{eq:local_fdr}. $\delta_j$ is a binary decision function defined as $\delta_j(t) = \mathbf{1}(\nu_j > t)$ for a given threshold $t$.\\
Furthermore, when the model $f$ and the parameter $\pi_0$ are correctly specified, by using the law of total expectation, we have 
\begin{align*}
    \text{FDR} = \text{E}_{D} [ \text{BFDR} ] = \text{E}_{D} \left[ \text{E}_{f} [ \text{FDP} \mid \text{D}] \right]
\end{align*}
i.e., on average, the frequentist FDR can be calculated as the expectation of the Bayesian FDR over repeated datasets.\\
The advantage of the \cite{Benjamini1995} procedure (BH) is that it does not rely on a specific modeling assumption on $f_1$ (the distribution under the alternative hypothesis) and just set $\pi_0 = 1$, so it assumes the worst case scenario, and then find the threshold with an adaptive data-driven procedure (\cite{Efron2008}). The class of Bayesian models defined above, on the other hand, can be quite sensitive to misspecification of $f_1$ and $\pi_0$.

\subsubsection{Bayesian Mirror Statistic}
Our proposal is to go beyond the use of Bayesian mixture models (such as the one in Equation \ref{eq:mixture_model}) and use a more general model specification that can impose shrinkage on the regression coefficients, without explicitly modeling and estimating inclusion parameters (akin to $\pi_0$ above). FDR control is then performed using the Mirror Statistic approach, calculated directly from the posterior distribution.\\
In this work we choose the same mirroring transformation used by \cite{Xing2021} and we approximate the distribution of $W_j$ through Monte Carlo draws as follows
\begin{align}\label{eq:MC_bayes_mirror_coef}
    w_j^{(s)} = m(\beta_j^{(s,1)}, \beta_j^{(s,2)} \mid D) = |\beta_j^{(s,1)} + \beta_j^{(s,2)}| - |\beta_j^{(s,1)} - \beta_j^{(s,2)}|
\end{align}
where $w_j^{(s)}$ represents a single value of $W_j$ and $\beta_j^{(s,1)}$ and $\beta_j^{(s,2)}$ are two independent draws from the posterior distribution $p(\beta_j | D)$. We repeat Equation \ref{eq:MC_bayes_mirror_coef} $N$ times for each coefficient $\beta_j$ to get a vector of values $\boldsymbol{w}_j = (w_j^{(1)}, \dots, w_j^{(s)}, \dots, w_j^{(N)})$.\\
This approach based on Monte Carlo approximation is generally applicable to every combination of posterior distribution and choice of $m(\cdot)$. However, for some combinations of the two, it might be possible to analytically compute the distribution of $W$. In Web Appendix A we provide an example of such a result, where we provide an analytical approximation to the distribution of $W$ when the posterior distribution of $\beta$ is Normal and when $m(\cdot)$ is defined as in Equation \ref{eq:MC_bayes_mirror_coef}. Once we have the Mirror Statistic samples $\boldsymbol{w}_j$, we can calculate the optimal inclusion threshold according to Equation \ref{eq:opt_t}, through which we calculate the inclusion probabilities for each covariate as:
\begin{align}\label{eq:inclusion_probs}
    \pi_j(t_{\alpha}) = \dfrac{1}{N} \sum_{s=1}^{N} \mathbf{1}\left( w_{j}^{(s)} > t_{\alpha} \right)
\end{align}
These probabilities play the same role as the local fdr defined in Equation \ref{eq:local_fdr}, but critically, these are not local fdr values, but rather joint measures of importance of all parameters $\beta_j$.\\
Finally, given these probabilities of inclusion, we can select the optimal subset of covariates for a given target FDR value $\alpha$ as follows
\begin{align}\label{eq:fdr_estimate_newton}
    \tau_{\alpha} &= \text{min}\left\{ \tau \in (0, 1): \text{FDP}(\tau) = \dfrac{\sum (1 - \pi_j) \mathbf{1}(\pi_j > \tau)}{\sum \mathbf{1}(\pi_j > \tau)} \leq \alpha \right\}
\end{align}
The quantity of interest in Equation \ref{eq:fdr_estimate_newton} is the numerator which works as an approximation of the number of false discoveries. Intuitively, if the classification is perfect and we get probabilities of inclusion that are nearly $1$ and $0$, respectively for true positive and true negative variables, then the approximation of the number of false discoveries will be exact. That is because the true positive coefficients will contribute almost zero to the sum, while each true negative coefficient will contribute with a weight which is close to $1$. The full procedure is summarised in Algorithm \ref{algo:bayesian_MS}.
\begin{algorithm}
\caption{Bayesian Mirror Statistic algorithm (\textit{BayesMS})}\label{algo:bayesian_MS}
\begin{algorithmic}[1]
    \Require FDR target value $\alpha$, posterior distribution $p(\beta | D)$. Then:\\

    Draw $N$ independent samples from the Mirror Statistic distribution $W$ (for example using Equation \ref{eq:MC_bayes_mirror_coef})\\
    Compute the optimal threshold $t_{\alpha} \in (0, \infty)$ using Equation \ref{eq:opt_t}\\
    Compute the inclusion probabilities $\pi_j(t_{\alpha})$ using Equation \ref{eq:inclusion_probs}\\
    Calculate the threshold $\tau_{\alpha}$ according to Equation \ref{eq:fdr_estimate_newton} and select the subset of covariates that satisfy $\pi_j(t_{\alpha}) > \tau_{\alpha}$
\end{algorithmic}
\end{algorithm}
We would like to highlight the flexibility of the proposed method in terms of adaptability to different regression problems. In contrast with alternatives, such as data splitting, model-X knockoff and Bayesian knockoff, our method does not require any major alteration of the model being used to perform inference. We only add an additional step of variable selection with FDR control on top of the output of the estimated model. This adds flexibility and the possibility to add the FDR step to existing models without required dedicated software packages.\\

In the following section we introduce some notation and we define the Bayesian models that we consider in this manuscript, as well as the inferential framework of choice. Then, in Section \ref{sec:simulations} we provide evidence of the performance of \textit{BayesMS} on complex simulated data, going through the full pipeline, from model specification to inference and variable selection.

\section{Model}\label{sec:model}
Let $f_{\boldsymbol{\theta}}(\boldsymbol{x}_i)$ be our model, where $f$ is the probability distribution representing the likelihood function, parametrized by $\boldsymbol{\theta}$. We complete the model specification by choosing an appropriate prior distribution for the high-dimensional inference target $\boldsymbol{\theta}$. As introduced in Section \ref{sec:Intro}, several priors have been proposed to perform inference in high-dimensional problems. A common choice is the Spike and Slab distribution, which explicitly provides posterior inclusion probabilities, but requires the exploration of a discrete space of mixture indicators which can be prohibitively slow in high-dimensions. In this work we avoid mixture priors by choosing fully continuous distributions, such as the Horseshoe prior of \cite{Carvalho2010}. Fully continuous distributions are computationally more efficient, however, they often do not provide explicit inclusion probability estimates like the Spike and Slab. In this work we use two continuous prior distributions, providing a way to effectively perform explicit variable selection, while also including FDR control using Algorithm \ref{algo:bayesian_MS}. We use the notation $\theta_j \sim \text{HS}(\sigma_{\tau})$ to refer to the Horseshoe prior, defined as:
\begin{align}\label{eq:horseshoe_prior}
\begin{split}
    \theta_j \mid \lambda_j, \tau &\sim \N(0, \lambda_j \tau)\\
    \lambda_j &\sim \text{C}^+(0, 1)\\
    \tau &\sim \text{C}^+(\sigma_{\tau})
\end{split}
\end{align}
where $\text{C}^+(\sigma_{\tau})$ refers to the Half-Cauchy distribution. This parametrisation has the desirable property of strongly shrinking null coefficients towards zero, while leaving non-null coefficients virtually unaffected. In our simulations we set $\sigma_{\tau} = 1$. Otherwise, $\tau$ can also be chosen a priori using cross-validation or other strategies based on a prior knowledge on the number of non-null coefficients (\cite{Piironen2017}).\\
An alternative prior specification, which we denote here as $\theta_j \sim \text{Prod}(a, b, \sigma_{\tau})$ is the following:
\begin{align}\label{eq:product_prior}
\begin{split}
    \theta_j \mid \eta_j, \lambda_j &= \eta_j \times \lambda_j\\
    \eta_j \mid \lambda_j, \tau &\sim \N(0, \lambda_j \tau)\\
    \lambda_j \mid a, b &\sim \text{Beta}(a, b)\\
    \tau &\sim \text{C}^+(\sigma_{\tau})
\end{split}
\end{align}
The idea behind this parametrisation is to shrink the effect $\theta_j$ by multiplying the coefficient by a proportion $\lambda_j$. Intuitively, if $\lambda_j \approx 0$, then the variance of $\eta_j$ will be small and at the same time the coefficient $\theta_j$ is pushed to zero. Viceversa, if $\lambda_j \approx 1$, the variance of $\eta_j$ will be approximately $\tau$ and $\theta_j$ is free to move away from zero. $\lambda_j$ enters at the same time in the prior specification and as a product element in the definition of $\theta_j$ in order to balance the two parameters.\\
We did not find a reference in the literature for this prior choice but we find it to be effective in performing shrinkage on the regression coefficients and also efficient in terms of optimisation. Moreover, it requires less tuning than the horseshoe prior, specifically, the critical shrinkage components $\lambda_j$ work well with a uniform prior, $\text{Beta}(1, 1)$, or with a symmetric prior choice, $\text{Beta}(0.5, 0.5)$. It could also be thought of as a completely continuous version of the Spike and Slab prior, which does not require a mixture distribution.

\subsection{Model inference}
The increased complexity of a Bayesian model comes with computational and methodological challenges. Markov Chain Monte Carlo (MCMC) methods are commonly used in Bayesian inference, but they are often inefficient in high-dimensional and more complex models. Variational Inference (VI) (\cite{Jordan1999, Blei2017}) is a computationally efficient alternative which casts the problem of integration into an optimization problem, thus opening the way to use modern gradient descent based algorithms and automatic differentiation.\\
VI looks for an approximation to the full posterior distribution $p(\boldsymbol{\theta} | y)$ by optimising the parameters of a family of distributions. The simplest (and most common) variational family is the factorised Normal distribution (also called Mean-Field or Isotropic Gaussian), where each parameter is approximated by an independent Normal distribution:
\begin{align}\label{eq:vi_MF}
\begin{split}
    q(\boldsymbol{\theta}) &= \prod_{k=1}^{|\theta|} q_{m_k, s_k}(\boldsymbol{\theta}_k)\\
    q_{m_k, s_k}(\boldsymbol{\theta}_k) &= \N (m_k, s_k)
\end{split}
\end{align}
where $|\theta|$ is the dimension of $\boldsymbol{\theta}$. This approximation is computationally very efficient but lacks the ability to capture dependencies in the posterior distribution. Starting from Equation \ref{eq:vi_MF}, we can increase the complexity by adding dependencies in the variational family, for example using a block structure to allow some group of parameters to be correlated. We choose to adopt this simple parametrization of the Variational distribution clearly because it is fast, but also because in such high-dimensional scenarios, estimating a full covariance matrix of the model coefficients might not be feasible at all. Moreover, although the covariance in the posterior is ignored, the weights $\boldsymbol{\theta}$, which are the actual target of the optimisation, are optimized all at once, accounting for the shape in the loss function dictated by the model.\\
Model inference is done through Automatic Differentiation Variational Inference (\cite{Kucukelbir2016}) and employing the Decayed Adaptive Gradient optimizer (\cite{Duchi2011}) which worked best for this class of models using the default parameters.\\

In the following section we provide extensive results based on diverse simulations. As the data generating mechanism becomes more complex, the model is also adapted.

\section{Simulations}\label{sec:simulations}
For our simulations we generate the covariates $\X$ from a multivariate Normal distribution, $\boldsymbol{x}_i \sim \N_p(\textbf{0}, \Sigma)$, where the covariance $\Sigma$ is constructed as a diagonal Toeplitz matrix, with each block defined as:
\begin{align}\label{eq:sigma_toeplitz}
    \begin{bmatrix}
        1 & \dfrac{(p^{\prime} - 2)\rho}{(p^{\prime} - 1)} & \dfrac{(p^{\prime} - 3)\rho}{(p^{\prime} - 1)} & \dots & \dfrac{\rho}{(p^{\prime} - 1)} & 0\\
        \dfrac{(p^{\prime} - 2)\rho}{(p^{\prime} - 1)} & 1 & \dfrac{(p^{\prime} - 2)\rho}{(p^{\prime} - 1)} & \dots & \dfrac{2\rho}{(p^{\prime} - 1)} & \dfrac{\rho}{(p^{\prime} - 1)}\\
        \vdots & & & \dots & & \vdots\\
        0 & \dfrac{\rho}{(p^{\prime} - 1)} & \dfrac{2\rho}{(p^{\prime} - 1)} & \dots & \dfrac{(p^{\prime} - 2)\rho}{(p^{\prime} - 1)} & 1
    \end{bmatrix}
\end{align}
where $p^{\prime}$ is the dimension of the block and $\rho$ (correlation factor) represents the highest value from which the correlation matrix is built.\\
The general strategy that we adopt is to run inference on 30 independent datasets and summarise the FDR and TPR (Power) calculated through Algorithm \ref{algo:bayesian_MS}. Where applicable, we compare the performance of our method against the \textit{knockoff}. Since we are dealing with simulated data, we know exactly the covariates covariance matrix and therefore we can use the \textit{knockoff} algorithm. The aim is to determine how well our algorithm can score compared with an ideal situation where the data generating mechanism is known, which is never the case with real data. Hence, we would expect that the knockoff method in practice will perform worse than this ideal scenario. We also tried to generate the knockoff variables using the Julia package \textit{Knockoffs.jl}, (\cite{Chu2024}), run with default settings, but the algorithm was not able to select any feature for the given simulated data.\\

\subsection{Linear model}
\begin{figure}
    \centerline{
    \includegraphics[width=1\linewidth]{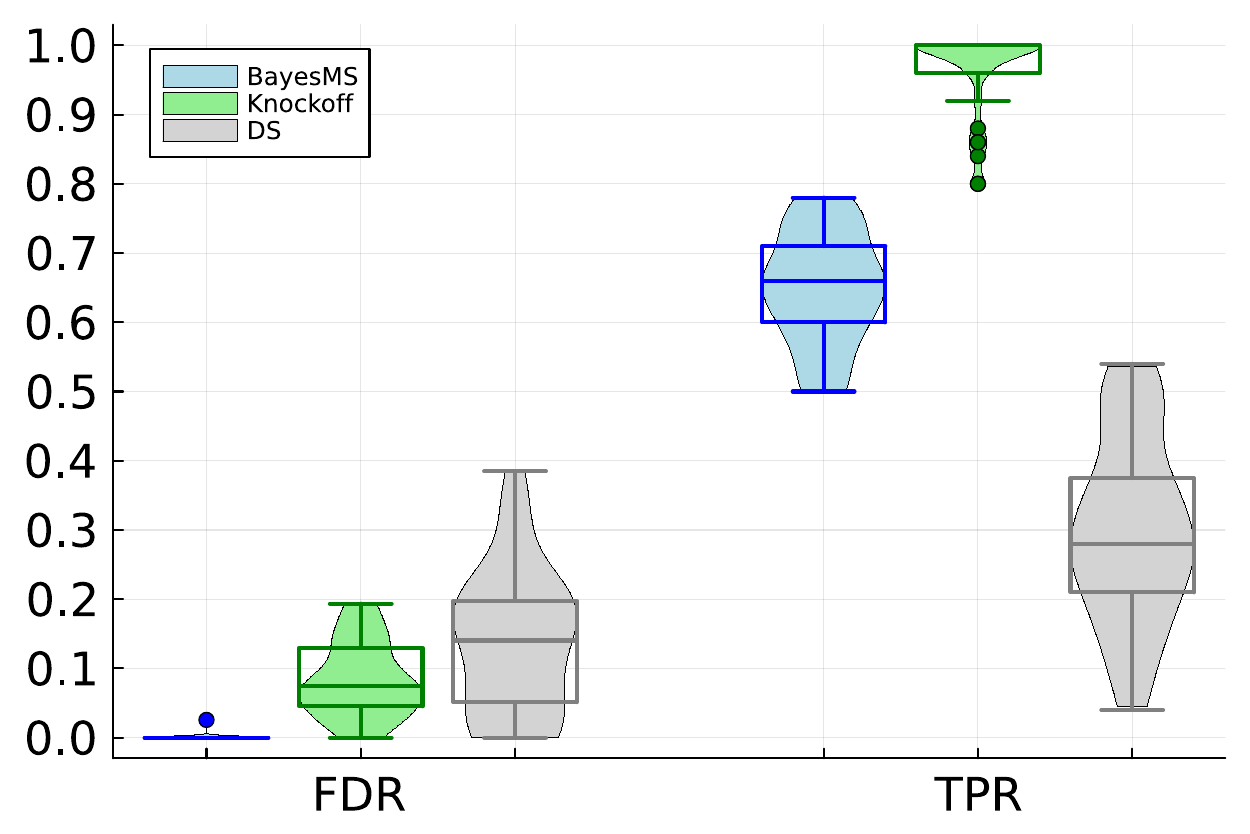}}
    \caption{Linear model (Equation \ref{eq:linear_model}) - FDR and TPR distributions for the variable selection on $\boldsymbol{\beta}$ obtained applying \textit{BayesMS}, compared with the \textit{knockoff} and the DS method}    \label{fig:algo_product_linear_n300_p1000_active50_r50_fdrtpr_boxplot}
\end{figure}

We first test \textit{BayesMS} FDR control strategy on a high-dimensional linear regression model. This is the only simulation where we can compare our method with the frequentist mirror statistic implementation via data splitting (\cite{Dai2022}).\\
The data is generated from the following process:
\begin{align}\label{eq:linear_model}
\begin{split}
    y_i = \xrow \boldsymbol{\beta} + \varepsilon_i\\
    \varepsilon_i \overset{IID}{\sim} \N(0, \sigma_{y})
\end{split}
\end{align}
The simulation framework consists of $n=300$ subjects, $p = 1000$ covariates, of which $p_0 = 950$ null coefficients and $p_1 = 50$ active coefficients. $\rho = 0.5$ and $p' = \{p_1, p_0\}$, respectively the correlation factor and block dimensions of $\Sigma$, the covariance matrix of $\X$. The non-null coefficients $\beta_j \in \{ -2, -1, 1, 2 \}$ and $\sigma_{y} = 1$.\\
For this simple linear regression problem we have tested both the horseshoe prior distribution (Equation \ref{eq:horseshoe_prior}) and the product prior distribution (Equation \ref{eq:product_prior}). Both choices work well in terms of variable selection and FDR control; here we show the results for the product prior. The model definition is as follows:
\begin{align*}
\begin{aligned}[t]
    y_i &\sim \N(\beta_0 + \xrow \boldsymbol{\beta}, \sigma_y) \\
    \beta_j &\sim \text{Prod}(1, 1, 1) \ \ \forall \ j=1,\dots,p
\end{aligned}
\qquad 
\begin{aligned}[t]
    \beta_0 &\sim \N(0, 5) \\
    \sigma_y &\sim \text{N}^+(1)
\end{aligned}
\end{align*}
In Figure \ref{fig:algo_product_linear_n300_p1000_active50_r50_fdrtpr_boxplot} we show the boxplots (overlapped with the violin plot) of the FDR and TPR for our proposed method, \textit{BayesMS}, together with two alternatives, the \textit{knockoff} method, which we can use since we know exactly the data generating mechanism, and the frequentist Mirror Statistic from data splitting (DS), which can also be used for high-dimensional linear models.
As expected, the knockoff method achieves, on average, FDR control around the target value of $0.1$, while also achieving a high TPR, with an average value close to 1. In this simulation we can see that our method is conservative in terms of FDR control, while maintaining a good TPR, with an average around $0.7$. For comparison we see how classic DS is not able to control FDR and also cannot achieve good levels of TPR.\\
The good performance of \textit{BayesMS} is partially due to the fact that the model can use the full dataset to perform variable selection and inference at the same time, as opposed to data splitting, an aspect which is critical in more extreme high-dimensional settings. The choice of the prior distribution also plays a role in effectively shrinking null coefficients to zero and automatically adapting to the sparseness level of the data.\\
In Web Appendix B we show for completeness the posterior distributions of the regression coefficients and the probabilities of inclusion.

\subsection{Linear random intercept model}
\begin{figure}
    \centering
    \includegraphics[width=0.8\linewidth]{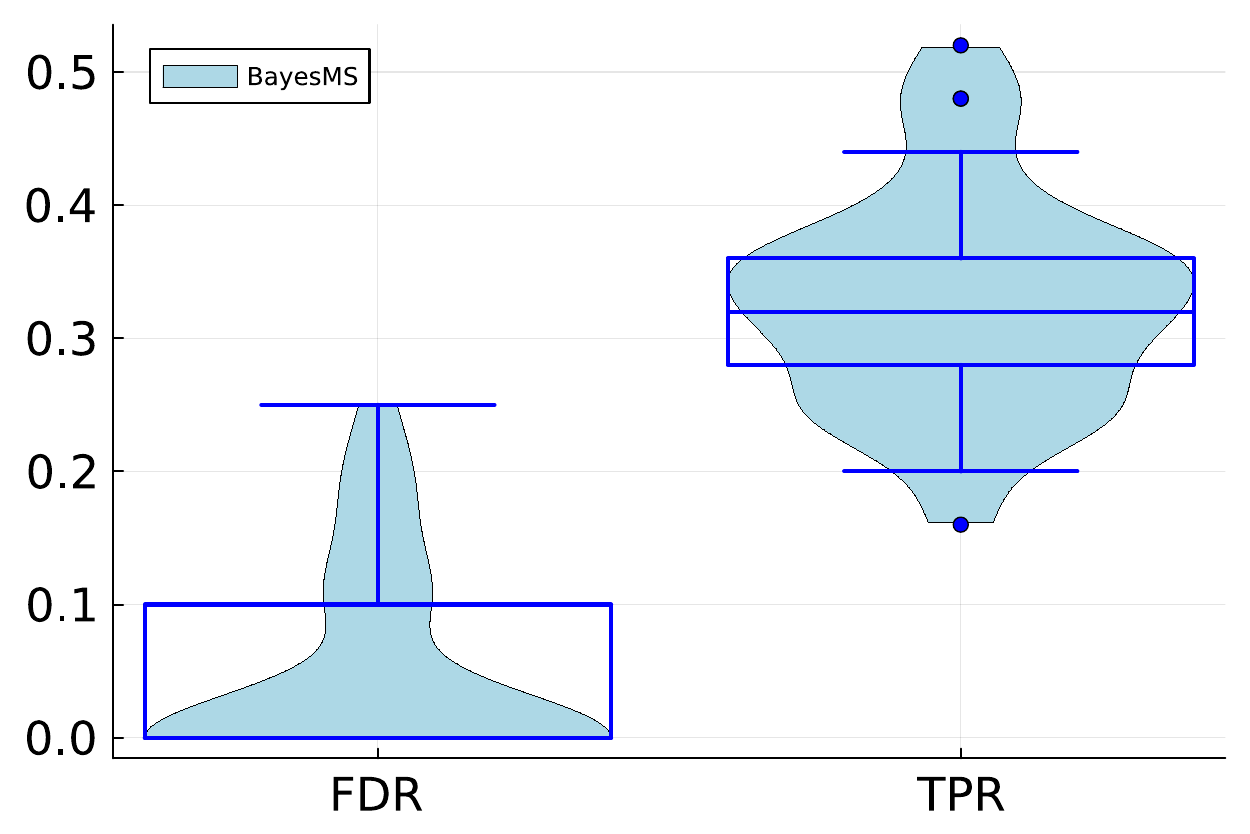}
    \caption{Random Intercept (Equation \ref{eq:random_intercept_model}) - FDR and TPR distributions for the variable selection on the fixed effects $\boldsymbol{\beta}$ (the median FDR is $0$)}
    \label{fig:algo_random_int_n100_M5_p500_active25_r50_fdrtpr_boxplot_R}
\end{figure}

We now extend the previous model to allow the analysis of repeated measurements. To this end we introduce a random intercept at baseline and we model the dependencies across multiple measurements through a hierarchical prior specification.\\
The data is generated from the following process:
\begin{align}\label{eq:random_intercept_model}
\begin{aligned}[t]
    &\mu_{il} = \beta_{0} + \beta_{0i}^R + \xrowr \boldsymbol{\beta}\\
    &\beta_{0i}^R \sim \N(0, \sigma_{\beta_0^R})
\end{aligned}
\qquad 
\begin{aligned}[t]
    &y_{il} = \mu_{il} + \varepsilon_{il}\\
    &\varepsilon_{il} \overset{IID}{\sim} \N(0, \sigma_y)
\end{aligned}
\end{align}
where the subscript $l = 1, \dots, M$, is the index of the \textit{l-th} repeated measurement and $\beta_{0i}^R$ is the random intercept.\\
This simulation framework consists of $n = 100$ subjects, $p = 500$ fixed effects, of which $p_0 = 475$ null coefficients and $p_1 = 25$ active coefficients, and $M = 5$ repeated measurements. $\rho = 0.5$ and $p' = \{p_1, p_0\}$, respectively the correlation factor and block dimensions of $\Sigma$. The non-null coefficients $\beta_j \in \{ -2, -1, 1, 2 \}$, $\sigma_{y} = 1$ and $\sigma_{\beta_0^R} = 2$.\\
We work with the following model specification
\begin{align*}
\begin{aligned}[t]
    y_i &\sim \N(\beta_0 + \beta_{i0}^{R} + \xrow \boldsymbol{\beta}, \sigma_y) \\
    \beta_j &\sim \text{Prod}(1, 1, 1) \ \ \forall \ j=1,\dots,p \\
    \beta_{i0}^{R} &\sim \text{N}(0, 3) \ \ \forall \ i=1, \dots, n
\end{aligned}
\qquad 
\begin{aligned}[t]
    \beta_0 &\sim \N(0, 5) \\
    \sigma_y &\sim \text{N}^+(1)
\end{aligned}
\end{align*}
The performance of \textit{BayesMS} is summarised in Figure \ref{fig:algo_random_int_n100_M5_p500_active25_r50_fdrtpr_boxplot_R}. FDR is properly controlled, with an average of about $0.05$, therefor being conservative. The power, expressed through the TPR, is on average around $0.35$, meaning that over half of the true active covariates have not been identified. However, we would like to highlight the fact that the smaller set of selected variables is not directly caused by the Mirror Statistic step, but rather by the model estimation itself. The posterior distributions have been shrunken towards zero, therefor leaving no evidence for the corresponding covariates to be selected.\\
From the simulations we notice that in high-dimensional settings, such as this one, it is increasingly difficult to properly estimate both the random intercept and the regression coefficients for each covariate. Note that in this particular example, we are not aware of any existing method to compare with.

\subsection{Generalised Linear models}
So far we have focused on variable selection in the presence of a continuous outcome. In this subsection we want to show the flexibility of our proposed method by controlling FDR in regressions with discrete outcomes, specifically distributed as Bernoulli and Poisson. This is an inherently more difficult task, but we show that we are still able to perform variable selection and FDR control.

\subsubsection{Logistic model}
\begin{figure}
    \centering
    \includegraphics[width=1\linewidth]{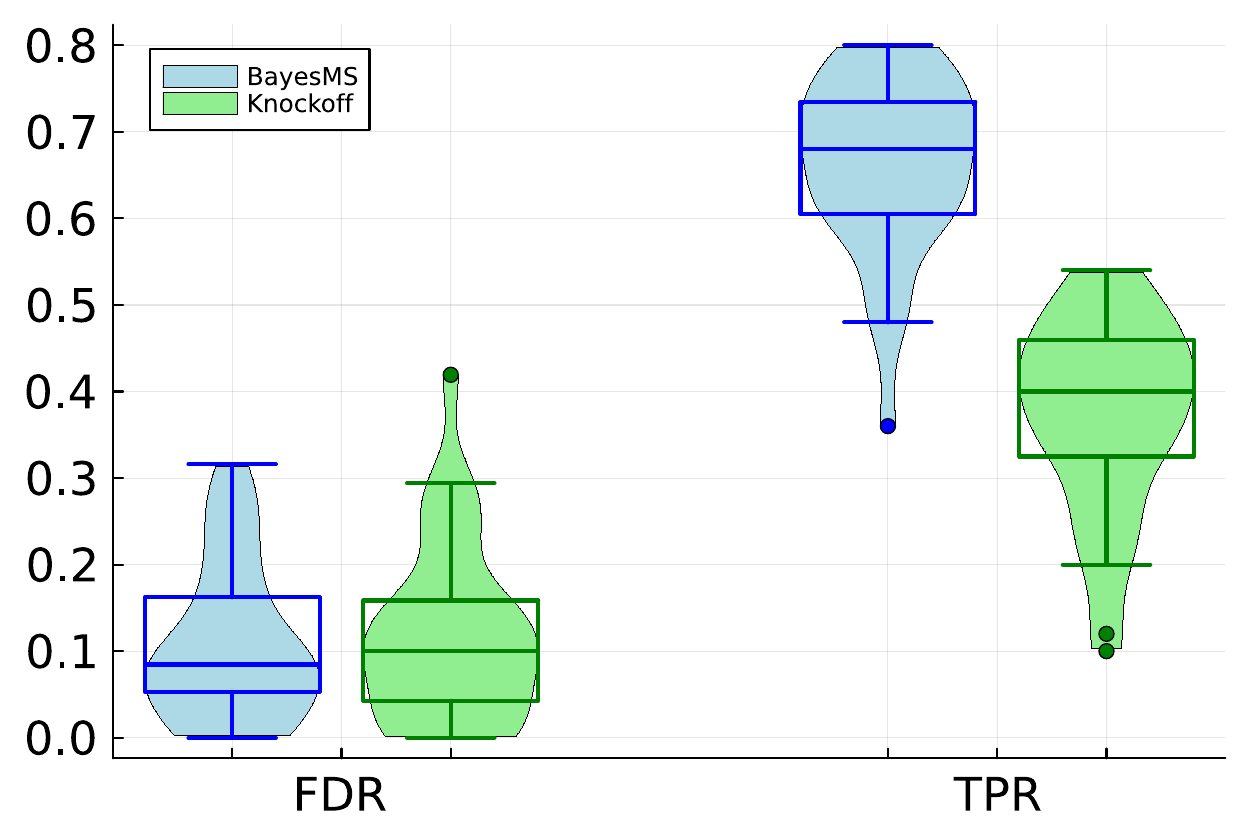}
    \caption{Logistic model (Equation \ref{eq:logistic_model}) - FDR and TPR distributions for the variable selection on $\boldsymbol{\beta}$ applying \textit{BayesMS} and the knockoff method.} \label{fig:algo_logistic_n500_p1000_active50_r50_fdrtpr_boxplot_R}
\end{figure}

For the Bernoulli distributed outcome we generate data from a high-dimensional logistic regression model as follows:
\begin{align}\label{eq:logistic_model}
\begin{split}
    \eta_i &= \xrow \boldsymbol{\beta}\\
    p_i &= logistic(\eta_i) = \dfrac{1}{1 + \exp{(-\eta_i)}}\\
    y_i &\overset{IID}{\sim} \text{Bernoulli}(p_i)
\end{split}
\end{align}
This simulation framework consists of $n = 500$ subjects, $p = 1000$ covariates, of which $p_0 = 950$ null coefficients and $p_1 = 50$ active coefficients. $\rho = 0.5$ and $p' = \{p_1, p_0\}$, respectively the correlation factor and block dimensions of $\Sigma$. The non-null coefficients $\beta_j \in \{ -2, -1, 1, 2 \}$.\\
For this experiment we have used the product prior distribution (Equation \ref{eq:product_prior}). The model is fully specified as
\begin{align*}
    y_i &\sim \text{Bernoulli}(logistic(\beta_0 + \xrow \boldsymbol{\beta})) \\
    \beta_j &\sim \text{Prod}(1, 1, 1) \ \ \forall \ j=1, \dots, p \\
    \beta_0 &\sim \N(0, 5)
\end{align*}
In Figure \ref{fig:algo_logistic_n500_p1000_active50_r50_fdrtpr_boxplot_R} we show the boxplots (and violin plots) of the Bayesian FDR and TPR, compared with the knockoff method, again in the idealized situation where we know the distribution of the covariates.
We can see how BayesMS is able to control FDR at the desired target level of $0.1$, while achieving an average TPR of almost $0.7$, highlighting the ability of the model to discover true signal, while limiting false discoveries. We also achieve a higher power compared to the theoretical knockoff.

\subsubsection{Poisson model}
\begin{figure}
    \centering
    \includegraphics[width=1\linewidth] {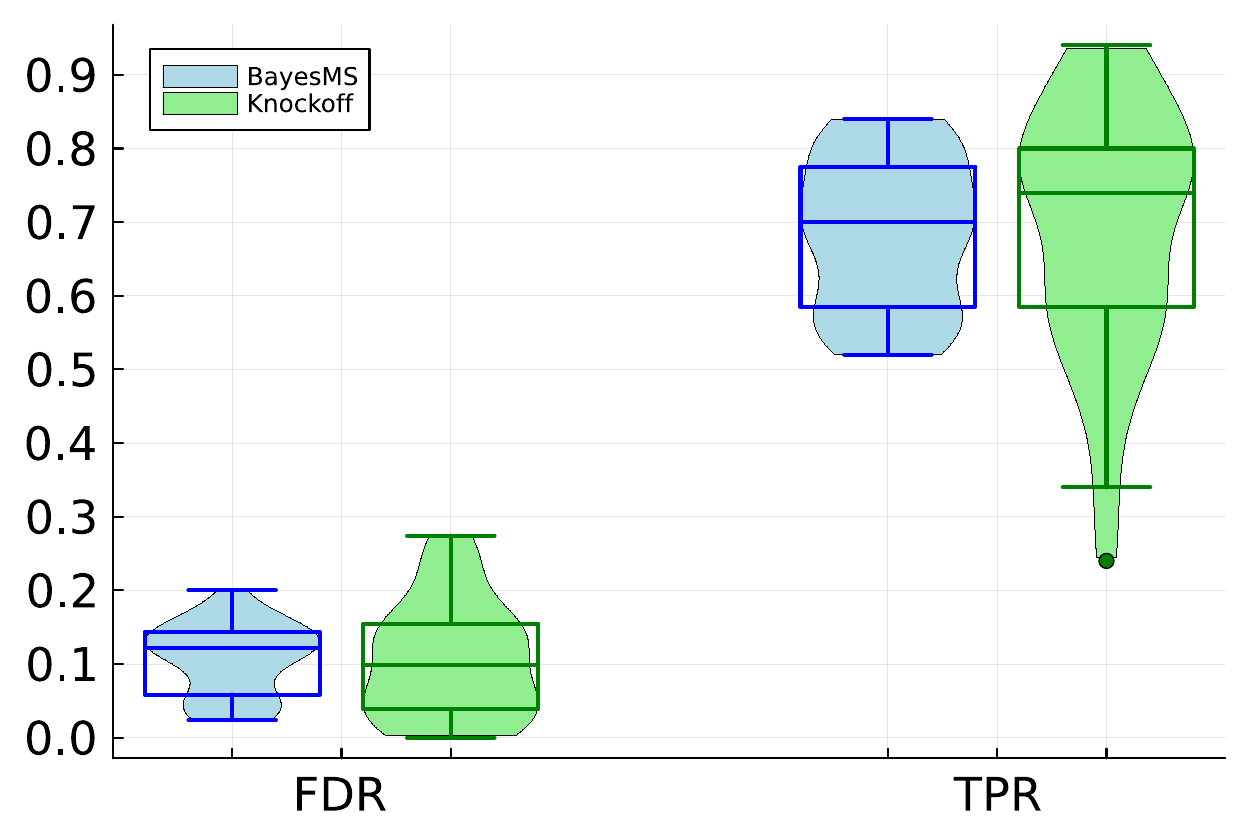}
    \caption{Poisson model (Equation \ref{eq:poisson_model}) - FDR and TPR distributions for the variable selection on $\boldsymbol{\beta}$ applying \textit{BayesMS} and the knockoff method.} \label{fig:algo_poisson_n500_p1000_active50_r50_fdrtpr_boxplot_R}
\end{figure}

In this additional simulation we want to highlight the flexibility of out proposed approach by estimating a high-dimensional model where the outcome is Poisson distributed. We generate data from a high-dimensional Poisson model as follows:
\begin{align}\label{eq:poisson_model}
\begin{split}
    \eta_i &= \beta_0 + \xrow \boldsymbol{\beta}\\
    \mu_i &= \exp{(\eta_i)}\\
    y_i &\overset{IID}{\sim} \text{Poisson}(\mu_i)
\end{split}
\end{align}
Here the simulation framework consists of $n = 500$ subjects, $p = 1000$ covariates, of which $p_0 = 950$ null coefficients and $p_1 = 50$ active coefficients. $\rho = 0.5$ and $p' = \{p_1, p_0\}$, respectively the correlation factor and block dimensions of $\Sigma$. The non-null coefficients $\beta_j \in \{ -1, 1 \}$ and $\beta_0 = 5$.\\
For this experiment we use the product prior distribution. The model is fully specified as
\begin{align*}
    y_i &\sim \text{Poisson}(\exp(\beta_0 + \xrow \boldsymbol{\beta})) \\
    \beta_j &\sim \text{Prod}(1, 1, 1) \ \ \forall \ j=1, \dots, p \\
    \beta_0 &\sim \N(0, 5)
\end{align*}
In Figure \ref{fig:algo_poisson_n500_p1000_active50_r50_fdrtpr_boxplot_R} we show the boxplots (and violin plots) of the Bayesian FDR and TPR, compared again with the knockoff method.
FDR is properly controlled by \textit{BayesMS}, which achieves an average at $0.1$. We are also able to achieve a good power, with an average TPR of about $0.7$. The idealized knockoff is also properly controlling the FDR and achieving a similar average TPR, but with a much higher variability in the performance.\\

These simulations provide evidence of the performance of \textit{BayesMS}, in particular the ability of the method to be used independently of the distribution of the outcome and/or the structure of the predictor.\\
Using Variational Inference, with a Gaussian variational family, we guarantee that the symmetry requirement is satisfied. This choice could be seen as a restriction since we are imposing the symmetry. However, we will argue that given the complexity of the problems that this work aims to solve, namely estimating coefficients in high-dimensional models with complex dependencies, this working assumption is reasonable and is shown to work well in practice.\\
In Supplementary Material we show additional results from the above simulations, as well as simulations on data generated from different models.

\section{Practical application}\label{sec:real_data}
Increased serum triglyceride (TG) level is a well-established risk factor for cardiovascular disease (CVD). Furthermore, the cardioprotective effects of marine omega-3 fatty acids are commonly attributed to their capacity to lower TG levels. Nevertheless, significant individual variability exists regarding TG responses to dietary fat intake. In this example, we will examine data derived from a randomized controlled crossover trial involving 47 healthy participants (both male and female), aged 25 to 46 years, with a mean body mass index of 23.6 kg/m² (\cite{Hansson2019}). The participants were provided with four different meals, each containing similar amounts of fat sourced from various dairy products. The TG responses were assessed through serum concentration measurements taken prior to the meal and at 2, 4, and 6 hours post-consumption. The primary objective of the original investigation was to evaluate the impact of these four different meals on TG responses. In addition to the principal exposure (the meal), we also have measurements of mRNA expression for a targeted set of genes and metabolomics data prior to each meal. Our primary focus lies in determining whether the TG response relates to mRNA levels at baseline, specifically exploring potential interactions between genes and time (in Web Appendix E we also analyse the metabolites concentrations). We apply our new method to investigate this problem and to identify potentially interesting genes.\\
Our focus is on determining whether baseline gene expression of the subjects affect the triglyceride trajectories over the 6 hours time interval. The number of covariates is $p = 625$, the gene expressions at baseline. This setting represents a challenging analysis as we need to deal with high-dimensional data, with repeated measurements over time and over different meals. Moreover, we would like to estimate potential effects of the covariates at baseline and their interaction with the time component, which represents one of the key factors of interest, therefore noticeably increasing the dimensionality of the problem.\\ 
We use the following model to analyse the data:
\begin{align*}
    &\mu_{ilt_0} = \beta_{0i}^R + \xrowr \boldsymbol{\beta}_{t_0} + \beta_l^{Meal} \ , \ \text{baseline}\\
    &\mu_{ilt} = \mu_{ilt-1} + \beta_{lt}^{Time} + \xrowr \boldsymbol{\beta}_{t}^{Inter}\\
    &y_{ilt} = \mu_{ilt} + \varepsilon_{ilt}\\
    &\varepsilon_{ilt} \overset{IID}{\sim} \N(0, \sigma_y)
\end{align*}
where the index $l$ represents the meal, $t$ represents the time and $i$ the subject. Here we assume that the regression coefficients $\boldsymbol{\beta}$ and the interaction coefficients $\boldsymbol{\beta}^{Inter}$ are shared across meals. This choice is supported by the clinicians we work with and it helps in reducing the dimensionality of the model.\\
We complete the model by specifying the following prior structure:
\begin{align*}
\begin{aligned}[t]
    \beta_l^{Meal} &\sim \text{N}(0, 1) \ \ \forall \ l \\
    \beta_{jt_0} &\sim \text{Prod}(1, 1, \tau) \ \ \forall \ j \\
    \tau &\sim \text{C}^+(1)\\
    \beta_{jt}^{Inter} &\sim \text{Prod}(1, 1, \tau_t) \ \ \forall \ j,t \\
    \tau_t &\sim \text{C}^+(1) \ \ \forall \ t
\end{aligned}
\qquad
\begin{aligned}[t]
    \beta_{0i}^{R} &\sim \text{HS}(1) \ \ \forall \ i\\
    \beta_{lt}^{Time} &\sim \text{N}(\mu_t^{Time}, \sigma^{Time}) \ \ \forall \ l=1,\dots,M \\
    \mu_t^{Time} &\sim \text{HS}(\sigma^{Time})\\
    \sigma^{Time} &\sim \text{C}^+(1)\\
    \sigma_y &\sim \text{N}^+(0, 0.5)
\end{aligned}
\end{align*}
For this analysis we fix the target FDR level at $20\%$, as the researchers see it as important to be able to pick up any potentially relevant findings. Running Algorithm \ref{algo:bayesian_MS} we are able to select $6$ coefficients, of which 4 at baseline (\textit{BTLS}, \textit{CCR1}, \textit{MCL1}, \textit{VTN}) and 2 interacting with time (\textit{CTLA4\_all} and \textit{C1QA}).\\
As expected this problem is particularly difficult, both because of the dimensionality and also because of the potentially very weak effect of gene expression on the outcome of interest. Nonetheless, we are able to identify some genes that might have an impact on the trajectory of triglyceride over time. Of particular interest are the two genes that interact with time (\textit{CTLA4\_all} and \textit{C1QA}), as these may play a role in regulation of triglyceride response.

\section{Discussion}\label{sec:Discussion}
In this paper we have proposed a new way of controlling FDR by adapting the frequentist Mirror Statistic into a Bayesian framework. In doing so we allow to perform FDR control for the regression coefficients of a wide range of models, such as linear regression and random coefficients models, but also logistic and Poisson regression, therefore allowing the analysis of both continuous and discrete outcomes.\\
The Mirror Statistic is a flexible approach that does not require the estimation of p-values. Likewise, our proposed approach does not require the estimation of inclusion probabilities, allowing a wide variety of models to be used. Moreover, we avoid the need of estimating two independent sets of coefficients by using the whole posterior distribution from the Bayesian model of choice. The only requirement is for the posterior distribution of interest to be symmetric and that the null coefficients are shrunk to zero. This behavior can be achieved with a wide choice of priors, with the Horseshoe and \textit{product prior}, which we have used, being two examples. This is in itself a great use of the inference provided by Bayesian inference, where the actual whole posterior distribution is used to run Algorithm \ref{algo:bayesian_MS}, rather then just a point summary.\\
We have tested the performance of the proposed method through extensive simulations, showing the ability to control the FDR at the desired target level, at most being overconservative in some cases. We compared the performance with the theoretical knockoff algorithm, i.e. the ideal scenario where the knockoff variables are generated knowing the full data generating process, a condition almost never available in practice. Only in the case of the linear regression we could also compare with the frequentist Mirror Statistic based on data splitting. For the high-dimensional random intercept model we did not find any ready available method for comparison.\\
An advantage of \textit{BayesMS} compared to other Bayesian FDR control strategies is the minimal changes required to perform FDR control; our method works directly on the posterior samples and does not require alteration of an existing model or data manipulation. The variable selection step is performed on top of the estimated model, thus the process of model estimation is independent of the FDR control. Lastly, the computational requirement are relatively low compared to Bayesian alternatives and comparable with frequentist models.\\
All simulations have been performed in \textit{Julia} (\cite{Bezanson2017}), version $1.11$. The code is freely available at \url{https://github.com/marcoelba/MirrorVI.jl}.\\

Several ways of improvement remain open, for example, providing a formal proof of the results that we have obtained through simulations. Another potential way of extending and improving the method would be to change the variational distribution. This can be done in several ways, for example by introducing dependencies across parameters using a multivariate Normal with a block covariance matrix. Another potential improvement could come from changing altogether the variational distribution, for instance rather than using a Gaussian, one could use a distribution that more closely resemble the parameter of interest. All these changes need, however, to be carefully evaluated against the additional computational cost and the increased complexity of the optimisation problem.

\newpage
\printbibliography

\section*{Code availability}
All simulations have been performed in \textit{Julia} (\cite{Bezanson2017}), version $1.11$, on Linux machine. The code is freely available at \url{https://github.com/marcoelba/MirrorVI.jl}.

\newpage
\section*{Appendix}
In this Appendix we provide more details about the Mirror Statistic distribution construction and we show additional results from the simulations.
 
\subsection{Mirror Coefficient distribution approximation}\label{sec:app_approximation_W}
In cases where the posterior distribution $p(\beta_j | D)$ is known and Normally distributed and the mirroring transformation $m(\cdot)$ is defined as in Equation \ref{eq:MC_bayes_mirror_coef}, we can further simplify the approximation of the distribution of $W$.\\
Given $\beta_j \sim \N(\mu, \sigma^2)$, the components in \ref{eq:MC_bayes_mirror_coef} are, respectively, the absolute value of the sum and the difference of two Normal distributions.\\
Given two independent Normal distributions, $X_1 \sim \N(\mu_1, \sigma_1^2)$, $X_2 \sim \N(\mu_2, \sigma_2^2)$, if $X = X_1 \pm X_2$, then $X \sim \N(\mu = \mu_1 \pm \mu_2, \sigma^2 = \sigma_1^2 + \sigma_2^2)$. Moreover, if $Y = |X|$, then $Y$ is distributed as a folded-Normal distribution, here denoted as $Y \sim \N^f(\mu_Y, \sigma_Y^2)$, with expectation and variance
\begin{align*}
\begin{split}
    \mu_Y = \text{E}(Y) &= \sqrt{\dfrac{2}{\pi}} \sigma \exp\left\{-\dfrac{2\mu}{\sigma^2}\right\} + \mu \left[ 1 - 2\Phi(-\dfrac{\mu}{\sigma}) \right]\\
    \sigma^2_Y = \text{Var}(Y) &= \mu^2 + \sigma^2 - \mu_Y^2
\end{split}
\end{align*}
So, according to Equation \ref{eq:MC_bayes_mirror_coef}, $w_j$ is defined as the difference between a folded Normal distribution with mean proportional to the mean of $\beta_j$ and a folded Normal distribution centred at $0$. Therefore, it is natural to construct $w_j$ as follows:
\begin{align}\label{eq:w_posterior_dist}
\begin{split}
    p(\beta_j | y) &= \N(\mu_{\beta_j}, \sigma^2_{\beta_j})\\
    p(w_j | y) &= \underbrace{ \N^f(\mu_{\beta_j}, \sigma^2_{\beta_j}) }_{A} - \underbrace{ \N^f(0, \sigma^2_{\beta_j}) }_{B}\\
    \implies p(w_j | y) &\approx \N(\mu_A - \mu_B, \sigma^2_A + \sigma^2_B)
\end{split}
\end{align}

\subsection{Simulation linear model}\label{sec:app_linear_model}
\begin{figure}[!ht]
    \centering
    \includegraphics[width=0.6\linewidth]{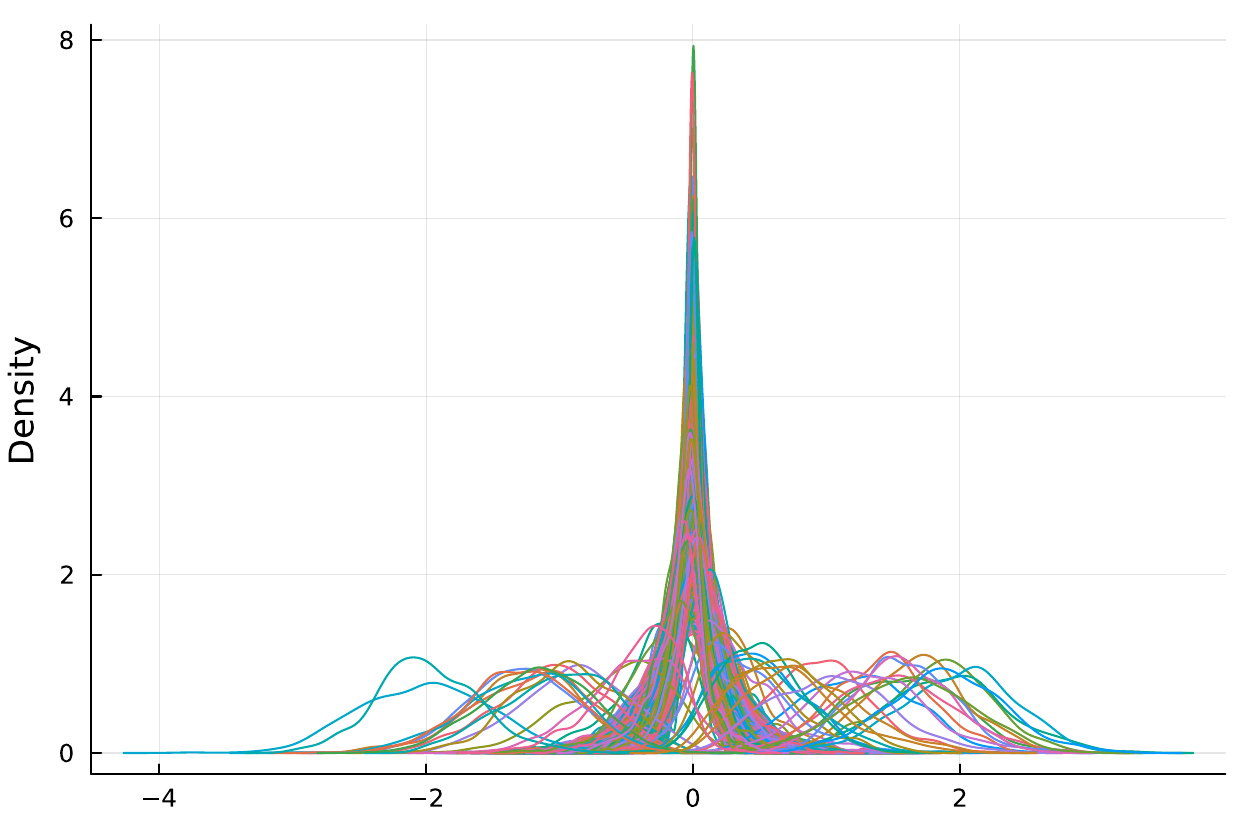}
    \caption{Linear Model (\ref{eq:linear_model}) - Posterior distribution of the regression coefficients $\boldsymbol{\beta}$}
    \label{fig:algo_product_linear_n300_p1000_active50_r50_posterior_beta}
\end{figure}

\begin{figure}[!ht]
    \centering
    \includegraphics[width=0.6\linewidth]{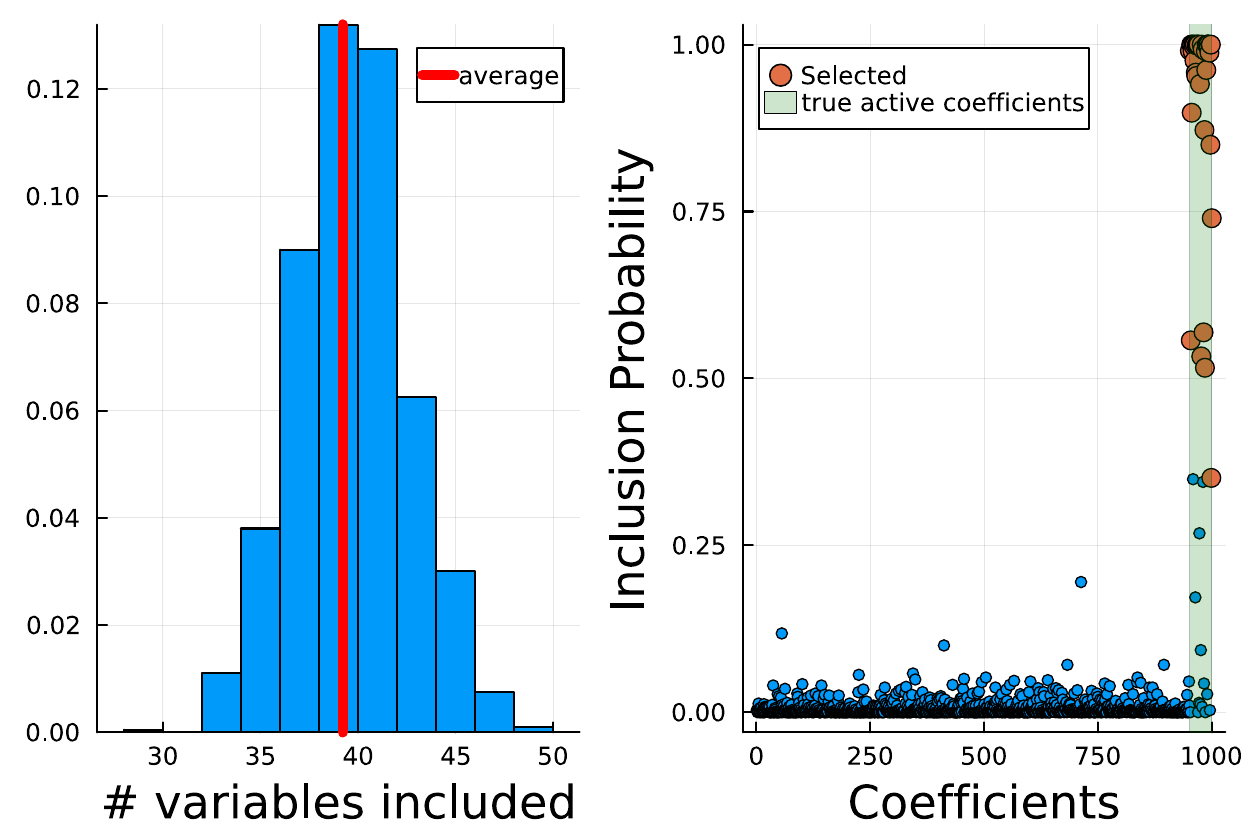}
    \caption{Linear Model (\ref{eq:linear_model}) - Left: Distribution of the number of variables included. Right: Inclusion probabilities and selected subset of covariates}
    \label{fig:algo_product_linear_n300_p1000_active50_r50_n_and_probs}
\end{figure}

Since we are working in a Bayesian framework, we can further analyse the posterior distributions to get additional insights on the parameters of interest. In Figure \ref{fig:algo_product_linear_n300_p1000_active50_r50_posterior_beta} we show the posterior distribution of the regression coefficients $\boldsymbol{\beta}$. We can appreciate how the shrinkage prior actively shrinks most coefficients to $0$, while still capturing the signal for some of the active coefficients. We can also see that the symmetry assumption required for the use of the Mirror Statistic is satisfied, as the distributions of the true null coefficients are symmetric around $0$.
In Figure \ref{fig:algo_product_linear_n300_p1000_active50_r50_n_and_probs} we show the distribution of the number of variables included (left) and the inclusion probabilities along with the selected covariates (right).

\subsection{Simulation logistic model}
\begin{figure}[!ht]
    \centering
    \includegraphics[width=0.6\linewidth]{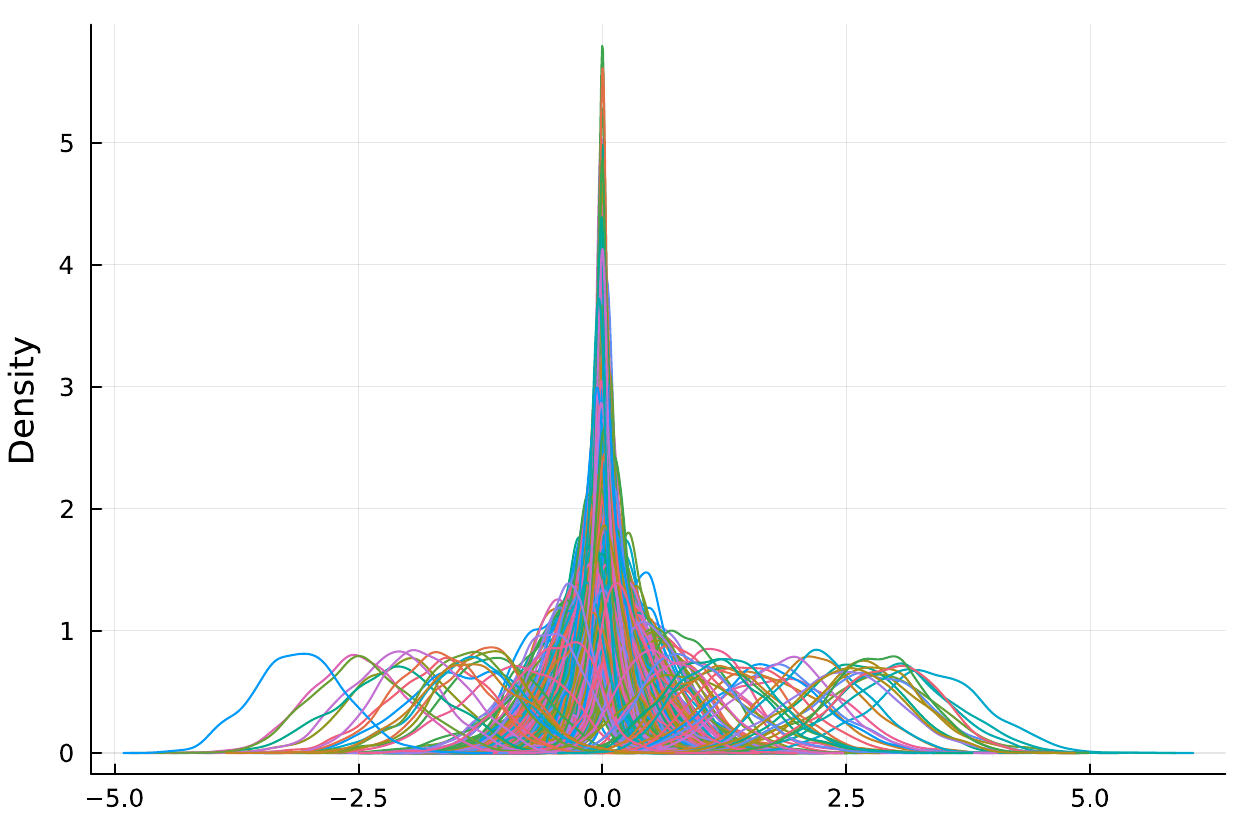}
    \caption{Posterior distribution of the regression coefficients $\boldsymbol{\beta}$}
    \label{fig:algo_logistic_n500_p1000_active50_r50_posterior_beta}
\end{figure}

\begin{figure}[!ht]
    \centering
    \includegraphics[width=0.6\linewidth]{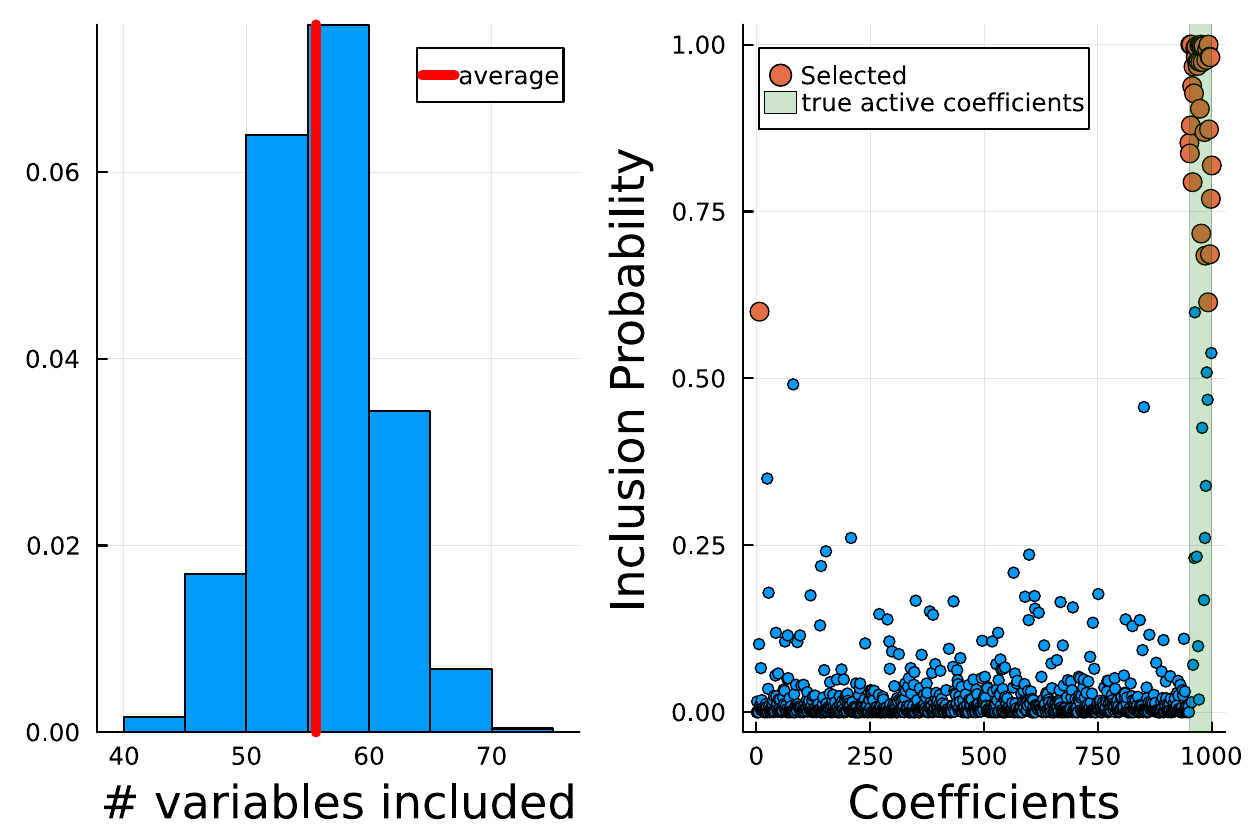}
    \caption{Left: Distribution of the number of variables included. Right: Inclusion probabilities and selected subset of covariates}
    \label{fig:algo_logistic_n500_p1000_active50_r50_n_and_probs}
\end{figure}

In Figure \ref{fig:algo_logistic_n500_p1000_active50_r50_posterior_beta} we show the posterior distribution of the regression coefficients $\boldsymbol{\beta}$. We are able to capture the signal for some of the active coefficients. We can also see that the symmetry assumption required for the use of the Mirror Statistic is still satisfied, even if the outcome is no longer Normally distributed.
In Figure \ref{fig:algo_logistic_n500_p1000_active50_r50_n_and_probs} we show the distribution of the number of variables included (left) and the inclusion probabilities along with the selected covariates (right).

\subsection{Simulation linear model with time dummies and repeated measurements}
In this experiment we introduce a time component into the data, plus repeated measurements at the patient level. This scenario is comparable with the real data analysis where we have multiple measurements represented by the meals, and multiple time points. To this end we introduce a random intercept at baseline and we model the dependencies across multiple measurements through a hierarchical prior specification. Data is generated from the following process:
\begin{align}\label{eq:time_interaction_model_repeated}
\begin{split}
    &\mu_{ilt_0} = \beta_{0} + \beta_{0i}^R + \xrowr \boldsymbol{\beta}_{lt_0} \ , \ \text{baseline}\\
    &\mu_{ilt} = \mu_{ilt-1} + \beta_{lt}^{Time} + \xrowr \boldsymbol{\beta}_{lt}^{Inter}\\
    &y_{ilt} = \mu_{ilt} + \varepsilon_{ilt}\\
    &\varepsilon_{ilt} \overset{IID}{\sim} \N(0, \sigma_y)
\end{split}
\end{align}
where the subscript $l = 1, \dots, M$, is the index of the \textit{l-th} repeated measurement, $\beta_{0i}^R$ is the random intercept, included only at baseline.\\
The model prior distributions are specified as follows: 
\begin{align*}
\begin{aligned}[t]
    \beta_{jlt_0} &\sim \text{Prod}(1, 1, \tau) \ \ \forall \ j, l \\
    \tau &\sim \text{C}^+(1)\\
    \beta_{jlt}^{Inter} &\sim \text{Prod}(1, 1, \tau_t) \ \ \forall \ j,l,t \\
    \tau_t &\sim \text{C}^+(1) \ \ \forall \ t
\end{aligned}
\qquad
\begin{aligned}[t]
    \beta_{0} &\sim \text{N}(0, 5)\\
    \beta_{0i}^{R} &\sim \text{HS}(1) \ \ \forall \ i\\
    \beta_{lt}^{Time} &\sim \text{N}(\mu_t^{Time}, \sigma^{Time}) \ \ \forall \ l\\
    \mu_t^{Time} &\sim \text{HS}(\sigma^{Time})\\
    \sigma^{Time} &\sim \text{C}^+(1)\\
    \sigma_y &\sim \text{N}^+(0, 0.5)
\end{aligned}
\end{align*}

For this experiment we use the following setting: 
\begin{itemize}
    \item $n = 100$ (sample size)
    \item $p = 100$ (fixed effects), $p_0 = 90$ (null coefficients), $p_1 = 10$ (active coefficients)
    \item $T = 4$ (time points)
    \item $M = 5$ (repeated measurements)
    \item $\rho = 0.5$, correlation factor for the block diagonal covariates correlation matrix
    \item $\beta_j \in \{ -2, -1, 0, 1, 2 \}$ for $j \in 1, \dots, p$
    \item $\boldsymbol{\beta}^{Time} = [ -2, -1, 0, 1, 2 ]$
    \item $\sigma_{\beta_0^R} = 5.$
    \item $\sigma_{y} = 1$, random error standard deviation
\end{itemize}

\begin{figure}[!ht]
    \centering
    \includegraphics[width=0.6\linewidth]{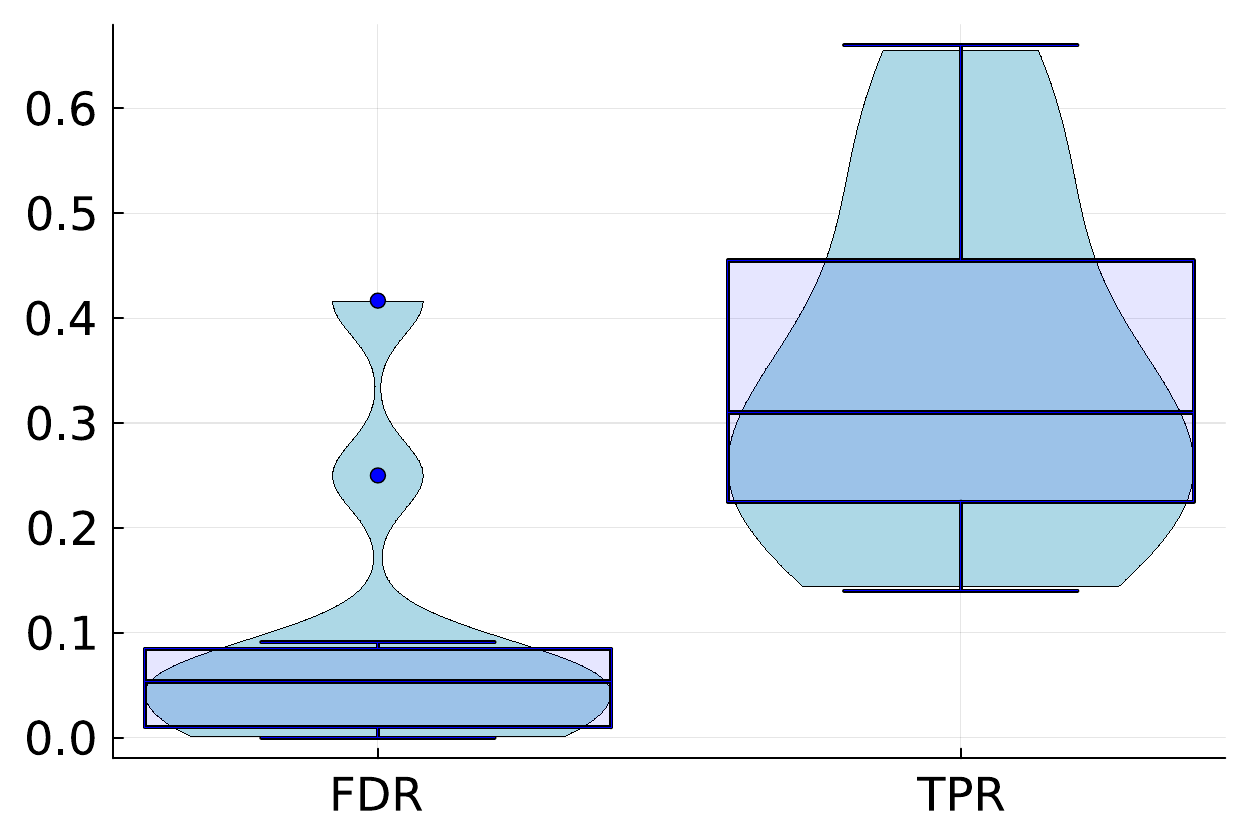}
    \caption{Linear time repeated measurements model (\ref{eq:time_interaction_model_repeated}) - FDR and TPR distributions for the variable selection on $\boldsymbol{\beta}$}
    \label{fig:algo_time_int_repeated_n100_t4_M5_p100_active10_r50_fdrtpr_boxplot}
\end{figure}

The performance is summarised in Figure \ref{fig:algo_time_int_repeated_n100_t4_M5_p100_active10_r50_fdrtpr_boxplot}
We can see that the FDR is on average below $0.1$, with two realisations over the target value. The TPR is lower compared to the case without repeated measurements, where we only had one set of regression coefficients, while here we let the coefficients be measurement specific. This leads to a substantial increase in the number of parameters to estimate, while the sample size increase is less due to the correlation between the observations.

\subsection{Metabolomics}
In addition to the gene expression data, as part of the nutrition longitudinal study, metabolomics data has also been collected. Being much lower in dimensionality compared to the genes, this is not an actual example of high-dimensional data, but the same model defined above can also be used to with meatabolites concentration instead of gene expression.\\
We have the same setting as we have with the genomics data, except the covariates, which are now of dimension $p = 37$, the metabolites measured at baseline. We are still using an FDR level of $20\%$.\\
We are able to identify 3 metabolites, \textit{MUFA} (Monounsaturated Fatty Acids) and \textit{Cit}, with an effect at baseline, and the \textit{ApoB\_ApoA1} ratio, which interact with time.

\end{document}